\documentclass[a4paper,11pt]{article}
\pdfoutput=1 % if your are submitting a pdflatex (i.e. if you have
\usepackage{jcappub} % for details on the use of the package, please
\usepackage[T1]{fontenc} % if needed

\title{\boldmath Possible Standard Model solution for Baryon Asymmetry}

\author[a,1]{M. Kosov,\note{Corresponding author.}}
\affiliation[a]{Dukhov Automatics Research Institute (VNIIA), Moscow, Russia}

\emailAdd{Kosov@vniia.ru}
\emailAdd{Mikhail.Kossov@jmail.com}

\abstract{If the discovered Higgs boson with $m_H$=125 GeV is interpreted as a  $t\bar{t}$-boson where the $t$-quarks are bound by Higgs-exchange with binding energy 220 GeV, then the $2t$-baryons should have approximately the same mass as the Higgs-boson. As $m_H<m_t$, the life time of $2t$-baryons must be much bigger then the life time of $1t$-baryons. If in the primordial Universe the number of $2\bar{t}$-antibaryons was bigger than the number of $2t$-baryons, then the excess should be compensated by nucleons. The  relatively long living heavy $2\bar{t}$-antibaryons could in primordial Universe fast evolve to antimatter black halls and disappear in the world of matter under the Schwarzschild spheres.}

\begin{document}
\maketitle
\flushbottom

\begin{figure}[tbp]
\centering
\includegraphics[width=0.80\columnwidth]{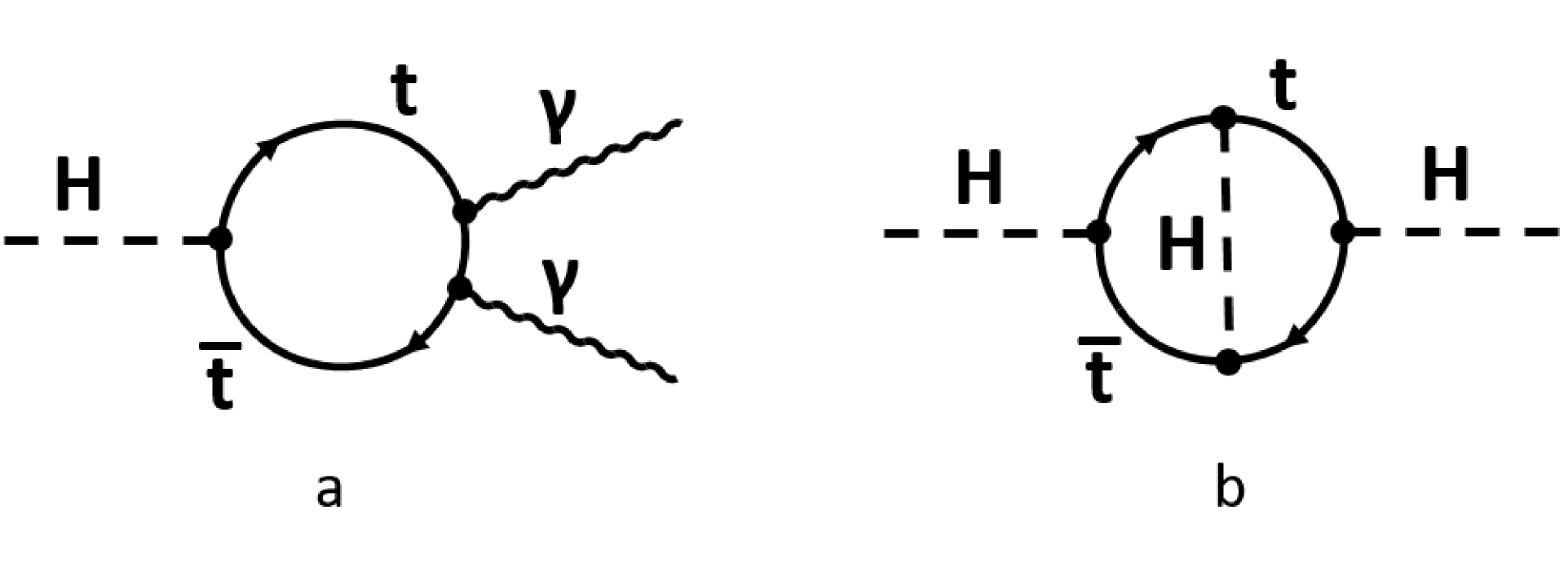}
\caption{\label{fig_Higgs} Higgs diagrams: (a) the $H\rightarrow\gamma\gamma$ decay, (b) the H self-interaction.}
\end{figure}

The fundamental enigma is "Where antimatter is hidden?”. Another philosophical enigma is "Why t-quarks exist, while they neither can be a part of any hadron, as their lifetime ($\approx 10^{-25}s$) is an order of magnitude smaller than the hadron formation time ($\approx 10^{-24}s$), nor, being colored, can exist by itself?”.
The third enigma is "How massive black holes could be created in Earlier Universe before nucleosynthesis started?”. These three enigmas could be solved by a set of low probable hypothesizes. The first hypothesis was discussed in ref.~\cite{csl:1}, where the discovered quantum of the Higgs field H-boson was interpreted as a $t\bar{t}$ pair bound by Higgs-exchange forces.
The Higgs boson was discovered in the $\gamma\gamma$-decay channel. The main diagram of the Higgs $\gamma\gamma$-decay includes the heaviest $t\bar{t}$ quark loop. It is shown in fig.~\ref{fig_Higgs}a.
The main feature of this diagram is the $tHt$-vertex, but the same $tHt$-vertex can define the  Higgs-exchange interaction inside the t-quark loop, as it is shown in fig.~\ref{fig_Higgs}b.
The second hypothesis is that the binding energy for the super-strong Higgs-exchange interaction, which reduces the 345 GeV total mass of the $t\bar{t}$ quark pair to the Higgs boson mass 125 GeV, is 220 GeV.
This hypothesis is questionable, but not impossible.

The Froggatt-Nielsen idea of the existence of the T-balls, consisting of 6 t-quarks and 6 $\bar{t}$-antiquarks, ref.~\cite{csl:2} was criticized in ref.~\cite{csl:3} and was discussed  even after the Higgs boson discovery in refs.~\cite{csl:4,csl:5}.
It means, that the $t\bar{t}$ binding energy can be at least comparable with the t-quark mass even in the Low Order approximation.
One can add another $t\bar{t}$ loops inside the diagram in fig.~{\ref{fig_Higgs}}b, and the resulting interaction can include more $t\bar{t}$ and $tt$ Higgs-exchange interactions making the binding energy big enough to reduce the mass of the $t\bar{t}$ quark pair and the mass of the $tt$-diquark to the Higgs boson mass.
It should be noted that the parity of the discovered Higgs boson is still not clear because of the big background.
\begin{figure}[tbp]
\centering
\includegraphics[width=0.60\columnwidth]{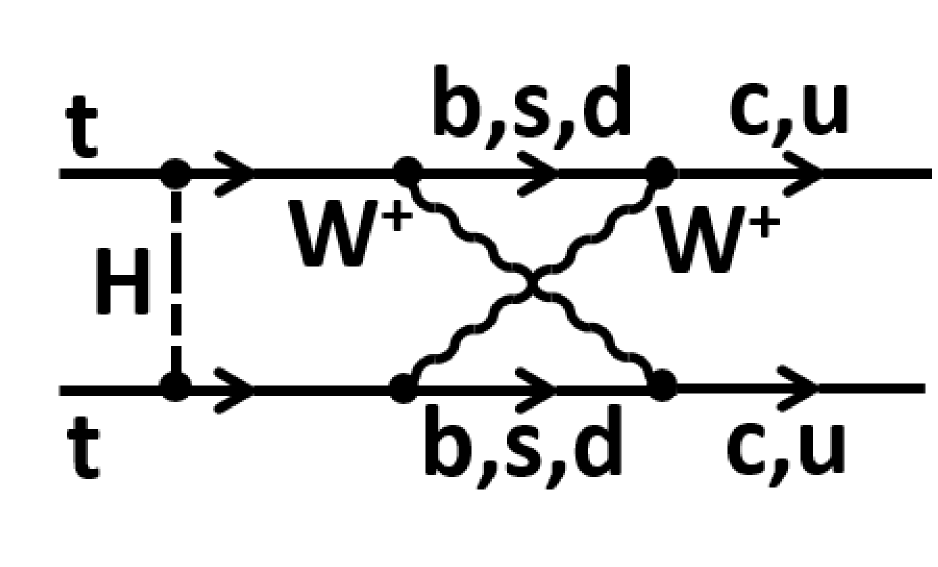}
\caption{\label{fig_ttuu} A diagram of the tt-diquark decay in the uu-diquark.}
\end{figure}

The hypothetical super-strong binding energy 220 GeV leads to the third hypothesis of the relatively long living 2t-baryons with $m_{tt}$ close to $m_H$.
The hypothetical 2t-baryons could be relatively long living because the decay of only one t-quark demands about 50 GeV of additional energy, as the 220 GeV binding is vanishing.
The hypothesis of the relatively long living 2t-baryon could be supported by the hard suppression of the $tt\rightarrow uu$ transition.
As the $(b,s,u)(b,s,u)$ intermediate states are forbidden by the charge conservation law, they must be compensated by the two W-boson cross-exchanges.
The possible diagram is shown in fig.~\ref{fig_ttuu}.
It would be very interesting to estimate the probability of such decay, but it is clear that it must be many orders of magnitude smaller than the probability of the $t\rightarrow (b,s,u)W$ decay in 1t-baryons.
The similar $\Xi\rightarrow N\pi$ decay is suppressed by at least 8 orders of magnitude.

If the $ttq$ and $tqq$ baryons together with the $\overline{ttq}$ and $\overline{tqq}$ antibaryons were created in the Big Bang so that the number of the $2\bar{t}$-antibaryons was bigger than the number of the $2t$-baryons, then after the fast decay of the $1t$-baryons and $1\bar{t}$-antibaryons the number of $2\bar{t}$-antibaryons must be bigger than the number of $2t$-baryons.
The excess of the heavy $2\bar{t}$-antibaryons should be compensated by nucleons according to the scheme: $3(dut)+(\overline{duu})+(\overline{dut})+(\overline{dtt})\rightarrow 3(duu)+2(\overline{duu})+(\overline{dtt})\rightarrow (duu)+(\overline{dtt})$.
Before inflation is finished the super-heavy $\overline{dtt}$-baryons could be collected by the gravitational forces in the primordial black holes and leave in the Universe only nucleons for the future nucleosynthesis.

As masses of the $2t$-baryons are two orders of magnitude bigger, than the nucleon masses, the evolution of the heavy 2$\bar{t}$-antibaryon stars to black halls during the inflation period could be rather fast, and the fourth hypothesis is that the time of the formation of these anti-black halls is comparable with the life time of the 2$\bar{t}$-antibaryons.
It looks impossible in our present Universe, but it can explain creation of the massive black halls in the earlier Universe ref.~\cite{csl:6}.

\begin{figure}[tbp]
\centering
\includegraphics[width=0.80\columnwidth]{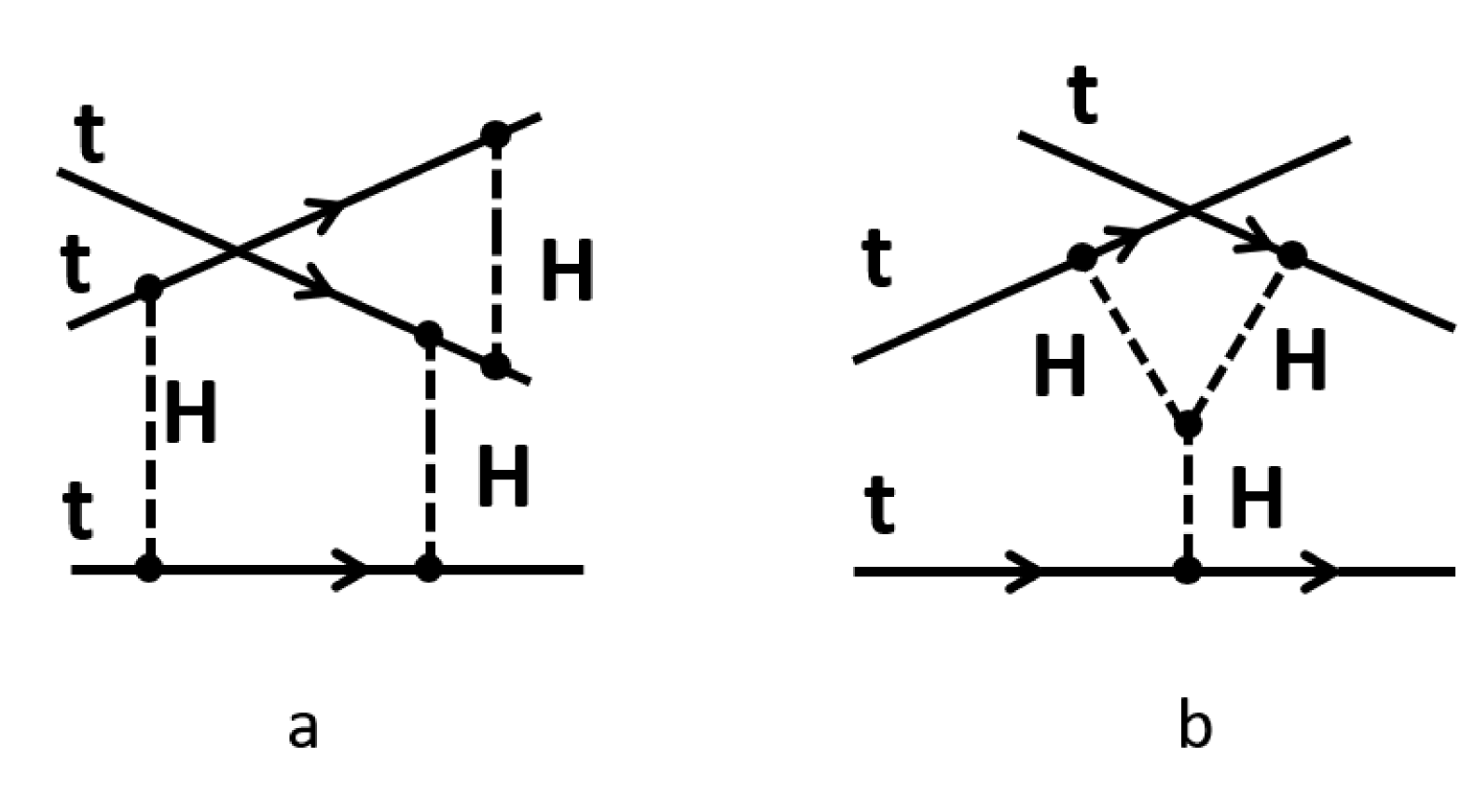}
\caption{\label{fig_3t} Diagrams of the bound 3t-baryon state.}
\end{figure}

The first hypothesis has a rather strange consequence mentioned in ref.~\cite{csl:1}. If the binding energy of each two t-quarks in fig.~\ref{fig_3t}a is 220 GeV, then the $3t$-baryon mass could be even negative (about -140 GeV).
But the binding energy can be reduced by the three particle interaction, which is shown in fig.~\ref{fig_3t}b.
The corresponding binding energy should be reduced by a factor $\frac{m_H}{m_t}$, and the $3t$-baryon mass could be positive (about 40 GeV).
As both the diagrams contribute to the 3t-baryon mass, it can have a small positive value.
If it is smaller than the nucleon mass, the $3t$-baryon must be stable and could be discovered experimentally.

In conclusion it should be stressed out, that our present knowledge can neither prove nor refute the proposed hypothesizes. The weakest point is the time of the evolution of stars made of the 125 GeV baryons. In the present Universe it takes a long time, but in the small volume of the Earlier Universe this process has a chance to be realized.


\begin{thebibliography}{99}

\bibitem{csl:1}
M. Kosov \emph{Masses of 92 1s-Hadrons in Chiral-Invariant Phase Space Model.} \textit{Phys. Atomic Nucl.} {\bf 88} (2025) 210-219.

\bibitem{csl:2}
C. D. Froggatt, H. B. Nielsen, L. V. Laperashvili \emph{Hierarchy-problem and a bound state of 6 t and 6 anti-t.} \emph{Int. J. Mod. Phys. A} {\bf 20} (2005) 1268

\bibitem{csl:3}
M. Yu. Kuchieve, V. V. Flambaum, E. V. Shuryak \emph{On bound states of multiple t-quarks due to Higgs exchange.} \emph{Phys. Rev. D} {\bf 78} (2008) 077502

\bibitem{csl:4}
M. De Santis \emph{Higgs interchange and bound states of super-heavy fermions.} \emph{Pramana} {\bf 81} (2013) 467.

\bibitem{csl:5}
C. D. Froggatt, C. R. Das, L. V. Laperashvili, H. B. Nielsen \emph{Diphoton decay of the Higgs boson and new bound states of top and anti-top quarks.} \emph{Int. J. Mod. Phys. A} {\bf 30} (2015) 1550132

\bibitem{csl:6}
I. Juodzbalis, R. Maiolino, W. M. Baker, \textit{et al.} \emph{A dormant overmassive black hole in the early Universe.} \emph{Nature} {\bf 636} (2024) 594-597

\end{thebibliography}
\end{document}